\documentclass[12pt]{article}
\usepackage{amsmath,amssymb}
\usepackage[colorlinks=true, pdfstartview=FitV, linkcolor=blue, citecolor=blue, urlcolor=blue]{hyperref}

\usepackage{color}

\newcommand{\ep}{\hfill  {\vrule height 10pt width 8pt depth 0pt}}

\newcommand{\bra}{\langle}
\newcommand{\ket}{\rangle}

\newcommand {\bT}{{\mathbb T}}
\newcommand {\bC}{{\mathbb C}}
\newcommand {\bZ}{{\mathbb Z}}
\newcommand {\bH}{{\mathbb H}}
\newcommand {\bI}{{\mathbb I}}

\newtheorem{thm}{Theorem} [section]
\newtheorem{lem}[thm]{Lemma}
\newtheorem{prop}[thm]{Proposition}
\newtheorem{defin}[thm]{Definition}
\newtheorem{cor}[thm]{Corollary}

\newtheorem {rem}[thm]{Remark}
\newtheorem {rems}[thm]{Remarks}

\title{Spectral Stability of Unitary Network Models \thanks{ Supported by
FONDECYT 1120786, Anillo ACT-1112,
and ECOS-Conicyt C10E10}}
\author{Joachim Asch  \thanks{Aix Marseille Universit\'e, CNRS, CPT UMR 7332, F--13288 Marseille cedex 9, France, e-mail : asch@cpt.univ-mrs.fr}\ 
\thanks{Universit\'e de Toulon, CNRS, CPT UMR 7332,  B.P. 20132, F--83957 La Garde, France},
Olivier Bourget
\thanks{
Departamento de Matem\'aticas
Pontificia Universidad Cat\'olica de Chile, Av. Vicu\~{n}a Mackenna 4860,
C.P. 690 44 11, Macul
Santiago, Chile},
Alain Joye
\thanks{
UJF-Grenoble 1, CNRS Institut Fourier UMR 5582, Grenoble, 38402, France}
}
\date{06.02.2015}
\begin{document}
\maketitle

\begin{abstract}
We review various unitary network models used in quantum computing, spectral analysis or condensed matter physics and establish relationships between them. We show that symmetric one dimensional quantum walks are universal, as are CMV matrices. We prove spectral stability and propagation properties for general asymptotically uniform models by means of unitary Mourre theory. 
\end{abstract}

\section{Introduction}

The last few years have witnessed a growing interest in several scientific communities for unitary network models defined on a lattice, or more generally on infinite graphs, describing the discrete dynamics of a quantum particle, possibly with internal degree of freedom. In condensed matter physics, popular models of this kind are the Chalker-Coddington model, \cite{cc, kok} describing the two-dimensional motion of electrons in a perpendicular magnetic field and a background potential and the Blatter and Browne model, \cite{bb}, accounting for the dynamics of electrons in a metallic ring subject to a constant electromotive force. In the field of theoretical quantum computing, the study of unitary models with internal degree freedom defined on various graphs, called generically quantum walks, are an active field of research. This is due to the instrumental role such dynamical systems play in the elaboration of quantum algorithms and efficiency tests of such algorithms, see e.g. \cite{ke, sa, gnvw, va}. Moreover, quantum walks also provide effective discrete models used in optics, be it to study atoms trapped in time periodic optical lattices, ions in suitably tuned magnetic traps, or polarized photons propagating  in networks of waveguides, \cite{Ketal, Zetal, scia}. Also, as the name suggests, quantum walks are sometimes considered as quantum analogs of classical random walks on the underlying graph, see {\it e.g.} \cite{ADZ, ke, sz, Ko, va}. This point of view has triggered interesting developments driven by analogies with classical probabilistic concepts,  \cite{Gu, APSS, skj, gvww, bgvw}. Last but not least, the celebrated 
CMV matrices associated with orthogonal polynomials with respect to a measure on the unit circle, and many of their extensions see \cite{S},  also belong to the class of unitary network models discussed in the present paper. In particular, their doubly infinite versions are closely linked to quantum walks, as made explicit in \cite{cgmv} for example.

This non exhaustive list illustrates the popularity of unitary network models and their flexibility in modelling various discrete unitary dynamical systems. Moreover, the algorithmic simplicity these models exhibit enables tractable, yet non trivial, mathematical analysis of their transport and spectral properties, which is the main focus of this paper. This trait of unitary network models has been exploited in the mathematical works mentioned so far and in \cite{bhj, dos, kos, msb, dfv, jma} for other examples of deterministic studies. For random versions of unitary models on cubic lattices or on trees see \cite{avww, joye, HJ} for temporal disorder, and \cite{HJS2, Ko1, jm, ASWe, ABJ1, ABJ2, j4, j1, HJ2}, for spatial disorder.\\

The present paper is devoted to the study of deterministic quantum unitary network models of the kind alluded to above. We first 
describe more precisely a few emblematic models on $\mathbb Z^d$ and establish some of their basic properties. We introduce in Section \ref{QW} the simple symmetric quantum walks on $\mathbb Z^d$ with internal degree of freedom in $\mathbb C^{2d}$, for an arbitrary dimension $d$. We  then discuss the Chalker-Coddington model on $\mathbb Z^2$ in Section \ref{seccc} and show that it can be written as a quantum walk. For $d=1$, we revisit the Blatter-Browne model, CMV matrices, and symmetric quantum walks and discuss their relationships in Section \ref{1dmodels}. There we show, see Remark \ref{rembbqw}, that unitary network models are generic in the sense that any unitary operator is unitarily equivalent to a quantum walk, extending a result of \cite{cgmv}. This provides further motivation to study  unitary network models. 

Then we consider generic deterministic quantum unitary network models defined as perturbations of translation invariant models, from a spectral perspective. More precisely, we are concerned with translation invariant models viewed as unitary matrix valued multiplication operators in the dual Fourier variable, i.e. fibered unitary operators, which typically exhibit purely absolutely continuous spectrum, a signature of transport. We consider multiplicative perturbations of such translation invariant models by operators that are multiplication operators in the discrete lattice variable, under certain assumptions on their behavior at infinity, as described in Section \ref{secmourre}. This setup corresponds to local perturbations of some homogeneous or periodic background in which the quantum particle propagates according to the discrete time dynamics induced by iteration of the quantum unitary model. Our results on the stability of spectral properties of translation invariant models perturbed that way are stated as Theorem \ref{mainmourre}.

The mathematical tool we use to get our spectral results is Mourre's method which is was originally developed to study stability of the  continuous spectrum of self-adjoint operators and was successfully applied to perturbations of self-adjoint translation invariant operators, see e.g \cite{m, abmg, gn} and references therein. Given our context, we need to resort to a unitary version of Mourre's theory, see Section \ref{gumt}, a topic under development in the recent years, see \cite{abcf, tfr, abc1}. 
Our main technical result regarding Mourre's unitary theory is described in Section \ref{secproof}. We define a self-adjoint conjugate operator associated with any translation invariant unitary network model that allows us to analyze the fairly general class of perturbed unitary operators loosely defined above. This class contains the main models discussed in the literature, and, in particular, those introduced in the first part of the paper. We end the paper by  spelling out the spectral consequences of our main abstract result on these examples in Section \ref{app}.

\section{Unitary Network Models}\label{uml}

\subsection{Quantum Walk on ${\mathbb Z}^d$}\label{QW}

We consider a simple symmetric quantum walk (QW for short) on ${\mathbb Z}^d$, with $2d$ complex internal degrees of freedom. It is customary to call ${\mathbb C}^{2d}$ the coin space.
Remark that  numerous variants of quantum walks exist in the literature,  each with its own merit, designed according to the context and goals considered. See for example \cite{hjs, hkss, va} and Section \ref{seccc}. We emphasize that our general result Theorem \ref{mainmourre} applies to many of these.

We now recall the  definition of a symmetric quantum walk. The canonical orthonormal bases of ${\mathbb R}^d$ and $l^2({\mathbb Z}^d)$ are denoted by  $\{f_k\}_{k=1,2,\dots, d}$, and $\{|j\ket\}_{j\in {\mathbb Z}^d}$ respectively.  We denote the canonical basis of  ${\mathbb C}^{2d}$,  by $\{|\tau\ket\}_{\tau\in {\cal N}_d}$, where the standard set of indices ${1, \dots, 2d}$ is relabelled as ${\cal N}_d= \{1, -1,\ldots, d, -d\}$. Following \cite{ke},  for example, we
introduce the following
\begin{defin}
\begin{itemize}
\item[i)] Let the symmetric shift operator $S$ on ${\mathbb C}^{2d}\otimes l^2({\mathbb Z}^d)$ be given by
\begin{equation}\label{convtau}
S=\sum_{j\in {\mathbb Z}^d} \sum_{\tau\in {\cal N}_d } |\tau \ket \bra \tau | \otimes |{j+\tau}\ket \bra j |,
\end{equation}
where we abuse notations by writing $j+\tau\in {\mathbb Z}^d$ to mean $j+\mbox{sign}(\tau)f_{|\tau|}\in {\mathbb Z}^d$.
\item[ii)]
For a given family of coin matrices $\{C(j)\}\in U(2d)^{{\mathbb Z}^d}$
the coin operator is defined as
\begin{equation*}
{\bf C}=\sum_{j\in {\mathbb Z}^d} C(j) \otimes | j\ket \bra j|\ \ \mbox{on} \ {\mathbb C}^{2d}\otimes l^2({\mathbb Z}^d).
\end{equation*}
\end{itemize}
The simple symmetric QW operator is then defined by the composition
\begin{equation*}
U=S {\bf C} \ \ \mbox{on} \ {\mathbb C}^{2d}\otimes l^2({\mathbb Z}^d).
\end{equation*} 
\end{defin}

The interpretation of $U$ is as follows: the action of $\bf C$ is local on the lattice and simply reshuffles the coin variables, whereas the action of $S$ makes the particle jump from its location on the lattice to its nearest  neighbors, according to the coin state.  
By construction $U$ only couples sites on the lattice that are at distance one apart,  consequently, after $n\in {\mathbb N}$ iteration, $U^n$ does not couple sites on the lattice that are a distance larger than $n$ apart. Remark that the evolution operator generated by a nearest neighbor hopping Hamiltonian is of infinite range in general.

In the constant case  $C(j)=C_\infty \in U(2d)$, for all $j\in {\mathbb Z}^d$, one speaks of a homogeneous QW and we denote by $U_\infty$ the corresponding QW operator. In this case $U_\infty$ is represented in Fourier space by a matrix valued multiplication operator by
\begin{equation}\label{mqw}
M(x)=\mbox{diag}(e^{ix_1}, e^{-ix_1}, e^{ix_2}, e^{-ix_2}, \dots, e^{ix_d}, e^{-ix_d})C_{\infty}, \ \ x\in \mathbb T^d.
\end{equation} 
More precisely with the Fourier transform defined on $L^2({\mathbb T}^d, dl; \mathbb C^{2d})$,  where $dl$ is the normalized Lebesgue measure, by
\begin{equation*}
{\cal F}: L^2({\mathbb T}^d, dl)  \rightarrow  l^2({\mathbb Z}^d)\qquad \ \ {\cal F}\left(e^{ij\cdot}\right): = |j\ket\qquad(j\in\bZ^d)
\end{equation*}
it holds:
\[\left(1\otimes{\cal F}\right)^{-1} U_\infty\left(1\otimes {\cal F}\right)\psi\otimes f(x)=\left(M(x)\psi\right)\otimes f(x)\qquad\left(\psi\in\bC^d, f\in L^2({\mathbb T}^d, dl) \right).\]
The matrix (\ref{mqw}) is the starting point of the abstract analysis performed in Section \ref{secmourre} below, where we shall  handle perturbations of this homogeneous situation. \\ 
\begin{rems} 
i) The method extends to  QW defined by a periodic configuration of coin matrices, at the price of increasing the dimension of the coin space, see \cite{AK1, bhj}. However, we shall not address this point here.\\
ii) The analysis of homogeneous quantum walks defined on trees is more subtle, due to the lack of an equivalent to the Fourier transform allowing us to express the QW operator as a multiplication operator, see \cite{jma}. 
\end{rems}

\subsection{Chalker-Coddington Model}\label{seccc}

The Chalker-Coddington effective model was introduced  in \cite{cc} in order to study the quantum Hall transition numerically in a quantitative way, see \cite{kok} for a review. Mathematical results on transport properties were given in \cite{ABJ1, ABJ2}. Our aim in this chapter is to prove that the model is equivalent to a simple symmetric QW.

The main features of the dynamics of a two-dimensional electron in a strong perpendicular magnetic field and a smooth bounded random potential are described by iterations of a random unitary ${U}$ acting on $l^{2}(\bZ^{2})$. The model (we refer to \cite{ABJ2} for more details) is  defined as  
\begin{equation*}
U(\varphi)=DT(\varphi)\qquad \hbox{ \rm on } l^{2}(\bZ^{2})\label{def:u}
\end{equation*}
where the matrix of the random unitary $D$ is diagonal in the canonical basis,  the angle $\varphi$ is a physical parameter and $T(\varphi)$ is the deterministic unitary operator
\[T(\varphi):=\cos\varphi S_{\circlearrowleft}+i \sin\varphi S_{\circlearrowright} \] 
built by superposition of local (anti-)clockwise rotations in the following sense: 

for the the canonical basis $\{ | j\ket\}_{j\in\bZ^{2}}$ of $l^{2}\left(\bZ^{2}\right)$, consider the decompositions
\[\bigoplus_{j\in\bZ^2}\bH_{\circlearrowleft}^{j}=l^{2}\left(\bZ^{2}\right)=\bigoplus_{j\in\bZ^2}\bH_{\circlearrowright}^{j}\]
where
\[\bH_{\circlearrowleft}^{j}:=\mbox{span}\left\{ |(2j_1,2j_2)\ket,  |(2j_1+1,2j_2)\ket,|(2j_1+1,2j_2+1)\ket,  |(2j_1,2j_2+1)\ket \right\},\]
\[\bH_{\circlearrowright}^{}:=\mbox{span}\left\{ |(2j_1,2j_2)\ket, |(2j_1,2j_2-1)\ket,|(2j_1-1,2j_2-1)\ket,  |(2j_1-1,2j_2)\ket \right\}.\]
Then
\[S_{\circlearrowleft}:=\bigoplus_{j\in\bZ^2} S_{\circlearrowleft}^{j}, \qquad S_{\circlearrowright}:=\bigoplus_{j\in\bZ^2} S_{\circlearrowright}^{j}\]

where for $\#\in\lbrace\circlearrowleft,\circlearrowright\rbrace$ the restrictions $S_{\#}^{j}$ of $S_{\#}$ to the invariant subspaces $\bH_{\#}^{j}$ are represented with respect to their basisvectors in the above indicated order by the permutation matrix
\[
\left(\begin{array}{cccc}
 0 & 0  & 0 & 1  \\
 1 & 0  & 0 & 0  \\
  0 & 1  & 0 & 0 \\
   0 & 0  & 1 & 0    \end{array}\right),\]
 
 i.e. $S_{\circlearrowleft}^{j}|2j_1,2j_2\ket=|2j_1+1,2j_2\ket\ldots$
\bigskip

We identify the Chalker Coddington model as a generalized quantum walk:

\begin{thm} For $\varphi\in\lbrack0,\frac{\pi}{2}\rbrack$, $U(\varphi)$ is unitarily equivalent to $\widetilde U(\varphi)$ on $\bC^4\otimes l^2(\bZ^2)$ defined by
\begin{equation*}
\widetilde U(\varphi)= {\bf D}(\cos\varphi  R\otimes\bI+i\sin\varphi (R^{-1}\otimes \bI) S)
\end{equation*}
where $R:\bC^4\to\bC^4$ is defined by
$R |\pm1\ket:=|{\pm2}\ket$, $R|\pm2\ket:=|\mp1\ket$ and
\begin{equation*}
S:=\sum_{j\in\bZ^2,\tau\in\{\pm1,\pm2\}}|\tau\ket\bra\tau |\otimes |j+\tau\ket\bra j|, \ \mbox{and} \ \ 
{\bf D}:=\sum_{j\in\bZ^2}D(j)\otimes |j\ket\bra j|
\end{equation*}
with the same convention as in (\ref{convtau}) regarding $j+\tau$, and $D(j)$  a diagonal unitary matrix.
\end{thm}
\begin{rem} Thus the Chalker Coddington model is a linear combination of symmetric quantum walks, one of them being static.
\end{rem}
{\bf Proof.}
Define the unitary operator ${\bf I}:l^2(\bZ^d)\to\bC^4\otimes l^2(\bZ^2)$ corresponding to the decomposition
$\bigoplus_{j\in\bZ^2}\bH_{\circlearrowleft}^{j}=l^{2}\left(\bZ^{2}\right)$ by
\[{\bf I}|2j\ket:=|-2\ket\otimes|j\ket,\quad
{\bf I}|2j+(1,0)\ket:=|+1\ket\otimes|j\ket\]
\[{\bf I}|2j+(1,1)\ket:=|+2\ket\otimes|j\ket,\quad
{\bf I}|2j+(0,1)\ket:=|-1\ket\otimes|j\ket.
\]
Then it clearly holds:  $\widetilde U(\varphi)={\bf I}U(\varphi){\bf I^{-1}}$.\ep

In Fourier space, ${\bf D}^{-1}\widetilde U(\varphi)$ is represented by the matrix valued multiplication operator by
\begin{equation}\label{mcc}
M(x_1, x_2)=\begin{pmatrix} 0 & i\sin(\varphi)e^{ix_2}& 0& \cos(\varphi) \\
\cos(\varphi) & 0 &  i\sin(\varphi)e^{-ix_1} &  0 \\
0&\cos(\varphi)& 0 &  i\sin(\varphi)e^{-ix_2} \\
 i\sin(\varphi)e^{ix_1} & 0 & \cos(\varphi) & 0 \end{pmatrix}.
\end{equation}

\subsection{QW, BB and CMV Models}\label{1dmodels}
In case the configuration space is one dimensional, in addition to the class QW described in Section \ref{QW}, we discuss two classes of unitary operators defined on $l^2({\mathbb Z})$, BB and CMV which have been considered in the physical or mathematical literature. 
BB operators appear as models in solid state physics, while CMV operators occur naturally in the study of orthogonal polynomials with respect to the unit circle, and therefore in the spectral analysis of general unitary operators. The goal in this section is to make explicit the relationships between the sets BB, CMV and QW.

\subsubsection{BB}
The set BB consists in two-sided infinite matrices defined on $l^2({\mathbb Z})$, with respect to the canonical basis, as a product of two operators, each of which given as an infinite direct sum of two by two unitary matrices, with matrix representations shifted by one. The name BB stands for Blattner and Browne who introduced these operators in \cite{bb} to study the dynamics of electrons in a metallic ring threaded by a time dependent magnetic flux.

With $P_{[j,j+1]}=|j\ket\bra j|+|{j+1}\ket\bra {j+1}|$, operators from BB are defined by 
\[U_{\text{BB}}=D_{\text{o}} D_{\text{e}} \hbox{ where} \]
\begin{equation}\label{bb}
 D_{\text{e}} = \sum_{k\in {\mathbb Z}}P_{[2k,2k+1]}S_{2k}P_{[2k,2k+1]},  \ D_{\text{o}} = \sum_{k\in {\mathbb Z}}P_{[2k+1,2k+2]}S_{2k+1}P_{[2k+1,2k+2]},
\end{equation}
and the unitary matrices $S_k\in U(2)$, called scattering matrices, are parametrized as
\begin{equation}\label{scatt}
S_k = e^{-i\theta_k}\left( \begin{array}{cc} r_k e^{-i\nu_k} & it_k e^{i\gamma_k}\\
it_k e^{-i\gamma_k} & r_k e^{i\nu_k}\end{array}\right), 
\hbox{ with}
\end{equation}
\[ (r_k,t_k)\in [0,1]^2, \quad r_k^2+t_k^2=1, \quad (\theta_k,\nu_k,\gamma_k)\in ({\mathbb T})^{3}\]
in the ordered basis $\{|k\ket, |{k+1}\ket\}.$
Explicitly, for any $k\in \mathbb Z$,
\begin{eqnarray}\label{matel}
U_{\text{BB}}e_{2k}&=&ir_{2k}t_{2k-1}e^{-i(\theta_{2k}+\theta_{2k-1})}
e^{-i(\nu_{2k}-\gamma_{2k-1})}|{2k-1}\ket\nonumber\\
&+&r_{2k}r_{2k-1}e^{-i(\theta_{2k}+\theta_{2k-1})}
e^{-i(\nu_{2k}-\nu_{2k-1})}|{2k}\ket\nonumber\\
&+&ir_{2k+1}t_{2k}e^{-i(\theta_{2k}+\theta_{2k+1})}
e^{-i(\gamma_{2k}+\nu_{2k+1})}|{2k+1}\ket\nonumber\\
&-&t_{2k}t_{2k+1}e^{-i(\theta_{2k}+\theta_{2k+1})}
e^{-i(\gamma_{2k}+\gamma_{2k+1})}|{2k+2}\ket\nonumber\\
& & \nonumber\\
U_{\text{BB}}e_{2k+1}&=&-t_{2k}t_{2k-1}e^{-i(\theta_{2k}+\theta_{2k-1})}
e^{i(\gamma_{2k}+\gamma_{2k-1})}|{2k-1}\ket\nonumber\\
&+&it_{2k}r_{2k-1}e^{-i(\theta_{2k}+\theta_{2k-1})}
e^{i(\gamma_{2k}+\nu_{2k-1})}|{2k}\ket\nonumber\\
&+&r_{2k}r_{2k+1}e^{-i(\theta_{2k}+\theta_{2k+1})}
e^{i(\nu_{2k}-\nu_{2k+1})}|{2k+1}\ket\nonumber\\
&+&ir_{2k}t_{2k+1}e^{-i(\theta_{2k}+\theta_{2k+1})}
e^{i(\nu_{2k}-\gamma_{2k+1})}|{2k+2}\ket.
\end{eqnarray}
Hence, all $U_{\text{BB}}$ have a five-diagonal band matrix 
structure
\begin{equation}\label{struct}
U_{\text{BB}}=\begin{pmatrix}\ddots & & & & & & & & &\cr
          \ast  &\ast &\ast &\ast & & & & & &\cr
          \ast  &\ast &\ast &\ast & & & & & &\cr
           & &\ast & \ast&\ast &\ast & & & &\cr
           & &\ast &\ast &\ast &\ast & & & &\cr
           & & & &\ast &\ast &\ast &\ast & &\cr
           & & & &\ast &\ast &\ast &\ast & &\cr 
           & & & & & &\ast &\ast &\ast &\ast \cr
           & & & & & &\ast &\ast &\ast&\ast  \cr
           & & & & & & & & &  \ddots           \end{pmatrix}   .
\end{equation}

When needed, we emphasize the dependence on the parameters in the notation by writing $U_{\text{BB}}(r, \theta, \nu, \gamma)$. Operators of this kind are studied in \cite{bhj}. See \cite{hjs} for a random version of BB operators and \cite{msb} for a generalization to similar operators constructed via higher dimensional scattering matrices.  Here we only   recall  some properties of BB operators in an informal way. The  phases $\{\gamma_k\}$ of $U_{\text{BB}}(r, \theta, \nu, \gamma)$  can be gauged away, see Lemma 3.2 in \cite{bhj}: let $V(\gamma)$ be defined by
\begin{equation}\label{unit}
 V(\gamma) |k\ket=e^{i\zeta_k}|k\ket,\,\,\,  k\in\mathbb Z,
\end{equation}
with $\zeta_0=0$ and
$
\zeta_k=-\sum_{j=0}^{k-1}\gamma_j, \,\, \zeta_{-k}=
\sum_{j=-1}^{-k}\gamma_j, \,\,  k\in {\mathbb N}^*.
$
Then, the following holds,
\begin{equation}\label{gauge}
V(\gamma)^{-1}U_{\text{BB}}(r, \theta, \nu, \gamma)V(\gamma)=U_{\text{BB}}(r, \theta, \nu, 0).
\end{equation}
Note that $V(\gamma+\tilde \gamma)=V(\gamma)V(\tilde\gamma)$, where $\gamma+\tilde \gamma$ is defined by componentwise addition in $\mathbb T^{\mathbb Z}$. \\
Finally,  if $S_{-1}$ is diagonal, the closed subspaces $\overline {\mbox{span}}\{|j\ket, j\in \mathbb N\}$ and \\ $\overline {\mbox{span}}\{|j\ket, -j\in \mathbb N^*\}$ are invariant and reduce $U_{\text{BB}}$, whereas if $S_0$ is diagonal, the closed subspaces $\overline {\mbox{span}}\{|j\ket, j\in \mathbb N^*\}$ and $\overline {\mbox{span}}\{|j\ket, -j\in \mathbb N\}$ are invariant and reduce $U_{\text{BB}}$. The corresponding  statements hold if $S_k$ is diagonal, for some arbitrary $k\in \mathbb Z$. 

We shall show in Lemma \ref{genericCMV} below that any unitary operator can be represented  by a direct sum of BB matrices of a special type thus BB matrices are universal.

\subsubsection{CMV}
By CMV we refer here to the set of doubly infinite five-diagonal matrices that extends the original definition of matrices on $l^2(\mathbb N)$ appearing naturally in the study of orthogonal polynomials on the unit circle and named after Cantero, Moral, Velazquez, \cite{cmv}. We recall here  a few facts about one sided unitary CMV matrices, referring the reader to \cite{S} for a detailed account on this topic which is the object of numerous investigations and extensions, in a deterministic and random framework.

One sided CMV matrices are to orthogonal polynomials  on the unit circle what Jacobi matrices are to orthogonal polynomials on the real axis: any unitary operator on a separable Hilbert space is given by a direct sum of one sided CMV matrices, which provide canonical models of cyclic unitary operators. 

Roughly speaking the construction goes as follows:  let  $U$ be a unitary operator on a separable Hilbert space $\cal H$. The spectral theorem says that $\cal H$ can be split into a finite or infinite direct sum of subspaces ${\cal H}_j$ generated by orthogonal vectors $\varphi_j \in \cal H$ that are cyclic for $U$. Moreover, $U|_{{\cal H}_j}$ is unitarily equivalent to the multiplication operator 
by $z$ in $L^2(\partial \mathbb D, d\mu_j)$, $d\mu_j$ being the spectral measure of the vector $\varphi_j$ and $\partial \mathbb D$ the unit circle.
One sided CMV matrices correspond to the multiplication operator by the independent variable $z\in\partial \mathbb D$ in $L^2(\partial \mathbb D, d\mu)$, expressed in a suitable orthogonal basis of Laurent polynomials in $z$, with respect to $d\mu$. 
They are characterized by an infinite sequence $\{a_k\}_{k\in {\mathbb N}}$, $a_k\in \mathbb D$, called Verblunski coefficients, defined by the construction of monic orthogonal polynomials with respect to $d\mu$. The Verblunski coefficients are in one to one correspondence with the measure $d\mu$ on $\partial \mathbb D$.
Remark that determining the one sided CMV form of a given cyclic operator is, however, not an easy task. 

Doubly infinite CMV matrices are denoted by $U_{\text{CMV}}$ and defined as special cases of BB matrices with  scattering matrices parametrized by Verblunski coefficients $\{a_k\}_{k\in {\mathbb Z}}$ given by
\begin{equation}\label{matcmv}
S_k=\begin{pmatrix}-|a_k|e^{i\mu_k}& \sqrt{1-|a_k|^2}\cr \sqrt{1-|a_k|^2}& |a_k|
e^{-i\mu_k}\end{pmatrix} =-i\begin{pmatrix}|a_k|e^{-i(\pi/2-\mu_k)}& i\sqrt{1-|a_k|^2}
\cr i\sqrt{1-|a_k|^2}& |a_k|e^{i(\pi/2-\mu_k)}
\end{pmatrix}
\end{equation}
where $a_k=|a_k|e^{i\mu_k}$. This corresponds to
the particular choices 
\begin{equation}\label{pcop}
\theta_k=\pi/2, \quad \nu_k=\pi/2-\mu_k, \quad r_k=|a_k|.
\end{equation}
The one sided CMV matrices $U^+_{\text{CMV}}$ with Verblunski coefficients $\{a_k\}_{k\in {\mathbb N}}$ discussed above are obtained by introducing boundary conditions at the site zero, choosing $S_{-1}=\mathbb I$, as
\begin{equation*}
U^+_{\text{CMV}}=U_{\text{CMV}}|_{\overline {\mbox{span}}\{|j\ket, j\in \mathbb N\}},
\end{equation*}
see Section 3 of \cite{cmv}, \cite{j} or Section 4 in \cite{S}.
 This way, any cyclic unitary operator can be
represented, in principle, by a doubly infinite CMV matrix, in the following sense:
\begin{lem}\label{genericCMV}
Let  $U$, a cyclic unitary operator on a separable Hilbert space $\cal H$, and $U^+_{\text{CMV}}$ be the corresponding one sided CMV matrix on $l^2(\mathbb N)=\overline{\mbox{span}}\{|j\ket, j\in \mathbb N\}$. Then, $U\oplus U$ on ${\cal H}\oplus {\cal H}$ satisfies
\begin{equation*}
\begin{pmatrix} U& \mathbb O \\ \mathbb O& U\end{pmatrix}\simeq \begin{pmatrix} U^-_{\text{CMV}}& \mathbb O \\ \mathbb O& U^+_{\text{CMV}}\end{pmatrix},
\end{equation*}
where $U^-_{\text{CMV}}$ on $\overline{\mbox{span}}\{|j\ket, -j\in \mathbb N^*\}$ is obtained by duplication of $U^+_{\text{CMV}}$:
\begin{equation*}\bra {-(j+1)}|U^-_{\text{CMV}} |{-(k+1)}\ket:=\bra {j}| U^+_{\text{CMV}}\ {k}\ket.
\end{equation*}
\end{lem}

\subsubsection{QW}
Finally, the set QW of simple one dimensional quantum walks acting on ${\mathbb C}^2\otimes l^2({\mathbb Z})$ described in Section \ref{QW} is characterized as follows. 
The coin operator is given by ${\bf C}=\sum_{j\in {\mathbb Z}^d} C(j) \otimes | j\ket \bra j|$, where, in the basis $\{|+1\ket, |-1\ket\}$ of $\mathbb C^2$,
\begin{equation*}
C(j)= e^{-i\eta_j}\left( \begin{array}{cc} \alpha_j & -\bar{\beta_j} \\ \beta_j & \bar{\alpha_j} \end{array} \right), \ \ \mbox{with 
$(\alpha_j,\beta_j)\in {\mathbb C}^2$ s.t $|\alpha_j|^2+|\beta_j|^2=1$, $\eta_j\in{\mathbb T}$.
}
\end{equation*}
The shift takes the form
$
S=\sum_{j\in {\mathbb Z}} |-1\ket \bra -1| \otimes |{j-1}\ket \bra j |+|+1\ket \bra +1| \otimes |{j+1}\ket \bra j |, 
$
whereas the corresponding quantum walk is denoted by $U_{\text{QW}}=S{\bf C}$. The dependence on the parameters will be denoted by $U_{\text{QW}}(\alpha, \beta, \eta)$.\\

We first note that the matrix representation of $U_{\text{QW}}(\alpha, \beta, \theta)$ in ${\mathbb C}^2\otimes l^2({\mathbb Z})\simeq l^2({\mathbb Z})$, takes the form of BB matrix, in a suitable basis. 
\begin{lem} A quantum walk  $U_{\text{QW}}(\alpha, \beta, \theta)$ is a BB matrix $U_{\text{BB}}(r, \theta, \nu, \gamma)$ with parameters given by (\ref{coeffbb}) below when expressed in the basis defined by
${\bf I}: {\mathbb C}^2\otimes l^2({\mathbb Z}) \rightarrow  l^2({\mathbb Z})$ s.t.
\begin{eqnarray*}
&&{{\bf I}} |+1\otimes k\ket = |{2k}\ket, \ \ 
{{\bf I}}|-1\otimes k\ket = |{2k+1}\ket.
\end{eqnarray*}
\end{lem}
{\bf Proof:} 
Explicit computations yield
\begin{equation*}
{{\bf I}}U_{\text{QW}}(\alpha, \beta, \eta){{\bf I}}^{-1}=U_{\text{BB}}(r, \theta, \nu, \gamma),
\end{equation*}
where the parameters $(r, \theta, \nu, \gamma)$ are determined by the scattering matrices,
\begin{equation}\label{coeffbb}
S_{2j+1}=i\begin{pmatrix}0&1\\ 1&0\end{pmatrix}, \ \mbox{and } \ S_{2j}=-i e^{-i\eta_j}\left( \begin{array}{cc}\beta_j & \bar{\alpha_j}  \\  \alpha_j & -\bar{\beta_j} \end{array} \right).
\end{equation}
\hfill \ep

The matrix representation of ${{\bf I}}U_{\text{QW}}(\alpha, \beta, \eta){{\bf I}}^{-1}$ is simpler than a generic BB matrix: 
\begin{equation}\label{structQW}
U_{\text{QW}}(\alpha, \beta, \eta)\simeq\begin{pmatrix}  & & & & &  \cr
          \ddots   &e^{-i\eta_{-1}}\bar\alpha_{-1}& & & &  \cr
           &0 & & & &  \cr
             & 0 & e^{-i\eta_{0}}\beta_{0} &e^{-i\eta_{0}}\bar\alpha_{0} & &  \cr
            &-e^{-i\eta_{-}}\bar\beta_{-1}&0 &0 & &  \cr
             & &0 &0 &e^{-i\eta_{1}}\beta_{1} & \cr
             & &e^{-i\eta_{0}}\alpha_{0}  &-e^{-i\eta_{0}}\bar\beta_{0}  &0  & \cr 
             & & & &0 &  \cr
             & & & &e^{-i\eta_{1}}\alpha_{1}   &\ddots   \cr
             & & & & &            \end{pmatrix}   .
\end{equation}

Conversely, to any BB matrix corresponds an explicit quantum walk operator that  represents the BB matrix in the sense of Proposition \ref{proot}. The argument is based on a parity symmetry that simple quantum walks possess.\\

Let  ${\mathbb C}^2\otimes l^2({\mathbb Z})={\cal L}_{\text{e}} \oplus {\cal L}_{\text{o}}$ where
\begin{equation}\label{subl}
{\cal L}_{\text{e}} =\overline{\mbox{span}} \{|\pm 1\otimes 2k\ket; k\in {\mathbb Z}\}, \ \ 
{\cal L}_{\text{o}} =\overline{\mbox{span}} \{|\pm 1\otimes 2k+1\ket; k\in {\mathbb Z}\},
\end{equation}
Both subspaces are identified with $l^2({\mathbb Z})$ via the unitary operators ${{\bf I}}_{\text{e/o}}: {\cal L}_{\text{e/o}}  \rightarrow  l^2({\mathbb Z})$ defined by:
\begin{eqnarray*}
&&{{\bf I}}_{\text{e}} |+1\otimes 2k\ket = |{2k}\ket, \ \ \ \ \ \ \ \ \ \ 
{{\bf I}}_{\text{e}} |-1\otimes 2k\ket = |{2k+1}\ket, \\
&&{{\bf I}}_{\text{o}}|+1\otimes 2k+1\ket = |{2k+1}\ket,\ \
{{\bf I}}_{\text{o}}|-1\otimes 2k+1\ket = |{2k+2}\ket.
\end{eqnarray*}
\begin{prop}\label{proot}
To any BB matrix $U_{\text{BB}}(r,\theta,\nu,\gamma)$ on $l^2(\mathbb Z)$ corresponds a QW operator $U_{\text{QW}}(\alpha,\beta,\eta)$ on ${\mathbb C}^2\otimes l^2({\mathbb Z})$ with  $\alpha_k=(-1)^kr_k e^{-i\nu_k}$, $\beta_k= it_ke^{-i\gamma_k}$ and $\eta_k=\theta_k$, $k\in\mathbb Z$, and a unitary operator $W: l^2({\mathbb Z})\oplus l^2({\mathbb Z})\rightarrow {\cal L}_{\text{e}} \oplus {\cal L}_{\text{o}}$ s.t. 
\begin{eqnarray}\label{sqrt}
&&U^2_{\text{QW}}(\alpha,\beta,\eta)=W\begin{pmatrix}U_{\text{BB}}(r,\theta,\nu,\gamma) & \mathbb O \\ \mathbb O & U_{\text{BB}}(r,\theta,\nu,\gamma)\end{pmatrix}W^{-1}, 
\end{eqnarray}
where $W={{\bf I}}^{-1}_{\text{\em e}} + {{\bf I}}^{-1}_{\text{\em o}}D^*_{\text{\em o}}(r,\theta,\nu,\tilde\gamma)V(\{\pi\})$, with $\tilde\gamma_k=\gamma_k+\pi$, $k\in\mathbb Z$.
\end{prop}
\begin{rems}\label{rembbqw}
i)  Any BB matrix, and {\it a fortiori} any CMV matrix can be described by a simple QW operator, in the sense of (\ref{sqrt}). Together with Lemma \ref{genericCMV}, it shows that any unitary cyclic operator $U$ can be represented by a simple quantum walk on ${\mathbb C}^2\otimes l^2(\mathbb Z)$, modulo multiplicity issues. This extends the statements of Section 7 in \cite{cgmv}.\\
ii) All propagation properties of the BB matrix are readily obtained from those of the corresponding QW operator since
\begin{equation*}
U^{2n}_{\text{QW}}=W \begin{pmatrix}U^n_{\text{BB}} & \mathbb O \\ \mathbb O & U^n_{\text{BB}}\end{pmatrix} W^{-1},
\ \  \forall\ n\in\mathbb Z.
\end{equation*}
\\
iii) By (\ref{gauge}) and $(-1)^j=e^{\pm i\pi j }$ it holds for the spectrum 
\[\sigma(U_{\text{BB}}(r,\theta,\nu,\gamma))=\sigma(U_{\text{BB}}(r,\theta,\nu,0))\hbox{ and }\] 
\begin{equation*}
\sigma(U_{\text{BB}}(r,\theta,\nu,\gamma))=\sigma(U^2_{\text{QW}}(\{r_je^{-i\nu_j}(-1)^j\},it,\theta))
=\sigma(U^2_{\text{QW}}(re^{-i\nu},it,\{\theta+j\pi\})).
\end{equation*}
\end{rems}
{\bf Proof: (of Proposition \ref{proot})} By construction, the operator $U_{\text{QW}}^2(\alpha,\beta,\theta)$ is reduced by the subspaces ${\cal L}_{\text{e}}$ and ${\cal L}_{\text{o}}$, so that if $P_{\text{e}}$ and $P_{\text{o}}$ denote the orthogonal projections on these subspaces, $U_{\text{QW}}^2(\alpha,\beta,\theta)= P_{\text{e}} U_{\text{QW}}^2(\alpha,\beta,\theta) P_{\text{e}} + P_{\text{o}} U_{\text{QW}}^2(\alpha,\beta,\theta) P_{\text{o}}$. One  checks  
that the first part of the decomposition yields the identity
\begin{equation*}
P_{\text{e}} U_{\text{QW}}^2(\alpha,\beta,\theta) P_{\text{e}}= D_{\text{o}}(r,\theta,\nu,\gamma)D_{\text{e}}(r,\theta,\nu,\gamma) , 
\end{equation*}
\mbox{with $\alpha_k=(-1)^kr_k e^{-i\nu_k}$, $\beta_k= it_ke^{-i\gamma_k}$}
which we use to fix the parameters of $U_{\text{QW}}^2(\alpha,\beta,\theta)$. The second part of the identity
yields 
\begin{equation*}
P_{\text{o}} U_{\text{QW}}^2(\alpha,\beta,\theta) P_{\text{o}}= D_{\text{e}}(r,\theta,\nu,\tilde\gamma)D_{\text{o}}(r,\theta,\nu,\tilde\gamma)\end{equation*}
\mbox{with $\alpha_k=(-1)^kr_k e^{-i\nu_k}$, $\beta_k= it_ke^{-i\tilde\gamma_k}$,}
where $\tilde \gamma_k =\gamma_k+\pi$, $k\in \mathbb Z$.
By definition, $U_{\text{BB}}=D_{\text{o}}D_{\text{e}}$ while $D_{\text{e}}D_{\text{o}}= D_{\text{o}}^*U_{\text{BB}} D_{\text{o}}$. Finally, property (\ref{gauge}) allows us to express $P_{\text{o}}U_{\text{QW}}^2(\alpha,\beta,\theta)P_{\text{o}}$ in terms of $U_{\text{BB}}(r,\theta,\nu,\gamma)$, with the initial parameters $\gamma$, which ends the proof.\ep

Thanks to Proposition \ref{proot}, we will formulate a spectral perturbation result for BB matrices in terms of the scattering matrices they are constructed from, see Corollary \ref{bbscat}, even though BB matrices do not have the structure assumed in Theorem \ref{mainmourre} below.

\section{Mourre Theory for Unitary Matrix Valued Multiplication Operators}\label{secmourre}
Our goal in this section is to establish a spectral stability result for unitary models represented by perturbed matrix valued multiplication operators.

In what follows, $d'\in {\mathbb N}$ and $M\in C^0({\mathbb T}^d; U(d'))$ where ${\mathbb T}^d$ (${\mathbb T}:={\mathbb R}/2\pi {\mathbb Z}$) is equipped with the normalized Lebesgue measure $dl$. We shall identify ${\mathbb T}$ and $\partial {\mathbb D}$, whenever convenient.
We abuse notations and denote by $M$ also the multiplication operator by $M(x)$ on $L^2({\mathbb T}^d; {\mathbb C}^{d'})$. Let  $\widehat{M}={\cal F} M {\cal F}^{-1}$ be the operator on $l^2({\mathbb Z}^d; \mathbb C^{d'})$ obtained by Fourier transform. The projections $p$ and $p_{\sigma}$ are defined on ${\mathbb T}\times {\mathbb T}^d$ by $p(\theta,x)=x$ and $p_{\sigma}(\theta,x)=\theta$. $$\Sigma:=\{(\theta, x) \in {\mathbb T}\times {\mathbb T}^d; \det (1 -e^{i\theta}M^*(x))=0\}$$
then $p_{\sigma}(\Sigma)=\sigma(M)=\sigma(\widehat{M})$.

We now consider an open set $\Theta\subset {\mathbb T}^d$ which avoids crossings and critical points of eigenvalues :
\begin{defin}\label{mgood} Given $M\in C^3(\mathbb T^d; U(d'))$, we say that an open set $\Delta\subset {\mathbb T}$ is $M$-good if there exists a finite family of disjoint open connected sets  $\{\Theta_j\}_{j=1}^N$ of ${\mathbb T}^d$, $N\in {\mathbb N}$, such that for $\Theta:= \bigcup_{j=1}^N \Theta_j$:
\begin{enumerate}
\item $\overline{p(p^{-1}_{\sigma}(\Delta) \cap \Sigma)}\subset \Theta$,
\item for any $j\in \{1,\ldots,N\}$ there are $k_j\in\left\{1,\ldots,  d'\right\}$ such that we have the spectral decomposition
\begin{equation*}
M(x)= \sum_{k=1}^{k_j} \lambda_{j,k}(x) \pi_{j,k}(x)\, , \quad \sum_{k=1}^{k_j} \pi_{j,k}(x) =1, \ \ \forall x\in \Theta_j, 
\end{equation*}
where  $\lambda_{j,k}(x)$, $\pi_{j,k}(x)$, 
are the eigenvalues, eigenprojections such that  $\lambda_{j,k}(x)\neq \lambda_{j,m}(x)$ if $k\neq m$ and 
$\nabla \lambda_{j,k}(x)\neq 0$.
\end{enumerate}
\end{defin}
\begin{rem}\label{rem:32}Under the conditions described in Definition \ref{mgood}, the maps $\lambda_{j,k}$ and $\pi_{j,k}$ are of class $C^3$ on $\Theta_j$. We also note that by definition, a $M$-good set $\Delta$ is a subset of $p_{\sigma}(\Sigma)$ so that $e^{i\Delta}\subset \sigma(M)=\sigma_{ess}(M)$.
\end{rem}

\begin{defin}\label{regular} A unitary operator on $l^2({\mathbb Z^d};  {\mathbb C^{d'}})\simeq {\mathbb C^{d'}}\otimes l^2({\mathbb Z^d})$ of the form ${\bf C}=\sum_{j\in \mathbb Z^d} C(j)\otimes |j\ket\bra j|$, with
$C(j) \in U(d')$ is called regular if
\begin{equation}\label{eq:regularity}
\int_1^{\infty} \sup_{a r\leq |j|\leq b r} \|C(j)-1\|\, dr < \infty \enspace ,
\end{equation}
for some $0<a<b<\infty$.
\end{defin}

\begin{thm}\label{mainmourre} Let $M\in C^3(\mathbb T^d ; U(d'))$, ${\bf C}$ be regular and $U:=\widehat M{\bf C}$. Let $\Delta$ be $M$-good. Then 
\begin{enumerate}
\item $\sigma_{sc}(U)\cap \Delta=\emptyset$,  $\sigma_{ac}(U)\cap \Delta=\sigma_{ac}(M)\cap\Delta$,
\item any compact set $\Delta'\subset \Delta$ contains only a finite number of discrete eigenvalues.
\end{enumerate}
If furthermore $M$ is analytic on $\mathbb T^d$, there exists a discrete set $\tau_M$ such that any open set $\Delta$ with $\overline{\Delta} \subset p_{\sigma}(\Sigma)\setminus \tau_M$ is $M$-good. It follows that $\sigma_{sc}(U)=\emptyset$.
\end{thm}
\begin{rem}  For ${\bf C}$ regular ${\bf C}-I$ is compact, thus $\sigma_{ess}(\widehat M)=\sigma_{ess}(U)$ by Weyl's Theorem. Hence, in gaps of $\sigma(\widehat{M})$, $U$ may only have discrete spectrum.\end{rem}

To prove Theorem \ref{mainmourre} we use unitary Mourre theory. We first review the essentials of the theory, then construct the relevant conjugate operator in case of unitary matrix valued multiplication. Note that under the same hypotheses, we get a limiting absorption principle, as explained below.

\subsection{General Unitary Mourre Theory}\label{gumt}

The regularity of an operator is defined via suitable commutation conditions w.r.t an auxiliary self-adjoint operator. In this section, ${\cal H}$ denotes a Hilbert space and $A$ a fixed self-adjoint operator  with domain ${\cal D}(A)$. For a unitary operator $U$, and  a Borel set of $\Delta\in \mathbb T$, we denote by $E_{\Delta}$ its spectral measure.

We define the class $C^1(A)$ as the family of bounded operators $B\in {\cal B}({\cal H})$ such that the sesquilinear form $Q$ defined on ${\cal D}(A)\times {\cal D}(A)$ by $Q(\varphi,\psi):=\langle A\varphi,B\psi\rangle - \langle \varphi,BA\psi\rangle$ is continuous w.r.t the topology induced by ${\cal H}\times {\cal H}$. The bounded operator associated to the extension of $Q$ to ${\cal H}\times {\cal H}$ is denoted by $\mathrm{ad}_A (B) =[A,B]$. 

We define $C^2(A)$  as the class of bounded operators $B\in C^{1}(A)$ such that $\mathrm{ad}_A B \in C^1(A)$. Equivalently, $B\in C^k(A)$ if and only if the map defined by $t\mapsto e^{-iAt}B e^{iAt}$ is strongly $C^k$, $k=1,2$. \cite{abmg}.
We also consider the following fractional order regularity.
\begin{defin}
A bounded operator $B$ is in ${\cal C}^{1,1}(A)$ if:
\begin{equation*}
\int_0^1 \|e^{iA\tau}B e^{-iA\tau}+e^{-iA\tau}B e^{iA\tau} -2B\|\,\frac{d\tau}{|\tau|^2} < \infty \enspace .
\end{equation*}
\end{defin}
\begin{rems} i) One has $C^2(A)\subset {\cal C}^{1,1}(A)\subset C^1(A)$. Furthermore ${\cal C}^{1,1}(A)$ is a $*$-algebra, see {\it e.g.}\ Section 5.1, \cite{abc1}. Mind that the integral is taken in the  norm sense.\\
ii) For unitary operators $U$ defined on ${\cal H}$, there exist alternative ways to show that $U\in C^1(A)$. Indeed, $U\in C^1(A)$ iff one of the following statements hold:
\begin{enumerate}
\item There exists a core for $A$, denoted ${\cal S}$, such that $U{\cal S} \subset {\cal S}$ and the sesquilinear form $F$ defined on ${\cal S}\times {\cal S}$ by $F(\varphi,\psi)=\bra U\varphi, AU\psi\ket -\bra\varphi, A\psi\ket$is continuous for the topology induced by ${\cal H}\times {\cal H}$.
\item There exists a core for $A$, denoted ${\cal S}$ such that $U{\cal S} \subset {\cal S}$ and the operator $U^*AU-A$ defined on ${\cal S}$ extends as a bounded operator on ${\cal H}$.
\end{enumerate}
iii) One can show that for $U\in C^1(A)$, the bounded operators given by the extensions of $F$ and $U^*AU-A$ coincide and are equal to $U^*\mathrm{ad}_A U$. See Section 6.2 of \cite{abc2}.
\end{rems}

Now, we introduce the concept of Mourre estimates for unitary operators:
\begin{defin}\label{propagating} Let $U\in C^1(A)$. For a given Borel set $\Delta \in {\mathbb T}$, we say that
$U$ is propagating with respect to $A$ on ${\Delta}$ if there exist $c >0$ and a compact operator $K$ such that: $E_{\Delta} (U^*AU-A) E_{\Delta} \geq c E_{\Delta} + K$. If $K=0$, $U$ is called strictly propagating. $A$ is called a conjugate operator for $U$.
\end{defin}
\begin{rem}\label{38} We observe that if $U$ is propagating w.r.t. $A$ on $\Delta$, then for any  $\phi\in C^0({\mathbb T}; {\mathbb R})$ supported on $\Delta$, $\phi(U) (U^*AU-A) \phi(U) \geq c \phi(U)^2 + K_{\phi}$ for some compact operator $K_{\phi}$.\\
 Conversely, if $\Delta$ is open and if for any $\phi\in C^0({\mathbb T}; {\mathbb R})$ supported on $\Delta$, $\phi(U) (U^*AU-A) \phi(U) \geq c \phi(U)^2 + K_{\phi}$ for some compact operator $K_{\phi}$, then for any Borel set $\Delta'$, such that $\overline{\Delta'} \subset\Delta$, we have $E_{\Delta'} (U^*AU-A) E_{\Delta'} \geq c E_{\Delta'} + K'$ for some compact operator $K'$.
 \end{rem}

{ We shall need the following result stated as Lemma 4.2 in \cite{abc2}:}
\begin{lem}\label{prop-compact} Let $U$ and $V$ be two unitary operators which belong to $C^1(A)$ and $c\in {\mathbb R}$.
\begin{enumerate}
\item If $U^*V-I$ and $\mathrm{ad}_A (U^*V)$ are compact, then given a real-valued function $\phi\in C^0({\mathbb T})$ we have that: $\Phi(U)(U^*AU-A)\Phi(U)\geq c\Phi(U)^2+K$ for some compact $K$ iff $\Phi(V)(V^*AV-A)\Phi(V)\geq c\Phi(V)^2+K'$ for some compact $K'$.
\item If $\mathrm{ad}_A (U^*V)$ is compact, then $(U^*AU-A)-cI$ is compact for some $c>0$ iff $(V^*AV-A)-cI$ is compact.
\end{enumerate}
\end{lem}
\begin{rem}  $U^*V-I$ is compact iff $U-V$ is compact. It this is the case and if we assume that the operators $U$ and $V$ belong to $C^1(A)$, then, $\mathrm{ad}_A (U^*V)$ is compact iff $\mathrm{ad}_A (V-U)$ is compact.
\end{rem}
We sum up the main results of unitary Mourre Theory in Proposition \ref{virial} and Theorem \ref{nosc}. By limiting absorption principle (LAP) for a unitary operator $U$ on some Borel subset $\Theta\subset {\mathbb T}$ w.r.t. a self-adjoint operator $A$, we mean:

\begin{itemize}
\item For any compact subset $\kappa\subset \Theta$
\begin{equation*}
\sup_{ |z| \neq 1, z\in \kappa} \|\bra A \ket^{-1}(1-zU^*)^{-1}\bra A \ket^{-1} \| < \infty \enspace .
\end{equation*}
\item If $z$ tends to $e^{i\theta} \in {\Theta}$ (non-tangentially), then $\bra A \ket^{-1}(1-zU^*)^{-1}\bra A \ket^{-1}$ converges in norm to a bounded operator denoted $R^+(\theta)$ (resp. $R^-(\theta) )$ if $|z| <1$ (resp. $|z| >1$). This convergence is uniform on any compact subset $\kappa\subset \Theta$.
\item The operator-valued functions defined by $R^{\pm}$ are continuous on each connected component of $\Theta$, with respect to the norm topology on ${\cal B}({\cal H})$.
\end{itemize}

\begin{prop}\label{virial} Assume that $U$ is propagating w.r.t $A$ on the Borel set $\Delta\subset {\mathbb T}$. Then, $U$ has a finite number of eigenvalues in ${\Delta}$. Each of these eigenvalues has finite multiplicity.
\end{prop}

\begin{thm}\label{nosc} Let $\Delta$ be an open subset of ${\mathbb T}$. Assume $U$ is propagating w.r.t. $A$ on $\Delta$ and, in addition,  $U\in {\cal C}^{1,1}(A)$. Then, a LAP holds for $U$ on ${\Delta} \setminus \sigma_{\text{pp}}(U)$ w.r.t. $A$. In particular, $U$ has no singular continuous spectrum in $\Delta$.
\end{thm}
See Section 4 of \cite{abc1} for the proofs of  \ref{virial} and \ref{nosc}.
\subsection{Proof of Theorem \ref{mainmourre}}\label{secproof}

With Definition \ref{mgood}, we set   $K_j:=\overline{p(p^{-1}_{\sigma}(\Lambda) \cap \Sigma)}\cap \Theta_j$, we fix $\eta_j\in C_0^\infty(\Theta_j)$ s.t. $\eta_j \upharpoonright K_j\equiv 1$, and consider $f_{j,k}:=i\lambda_{j,k} \nabla \overline{\lambda_{j,k}}=-i \overline{\lambda_{j,k}}\nabla \lambda_{j,k}$,  all $j\in \{1, \ldots,N\}, k\in \{1, \ldots, k_j\}$. Note that $f_{j,k}\in C^2(\Theta_j;{\mathbb R}^d)$. We define the symmetric operator $A_{M,\eta}$ on $C^{\infty}({\mathbb T}^d)$:
\begin{equation}\label{conjugate2}
A_{M,\eta}:= \frac{1}{2} \left[ \sum_{j=1}^{N}\eta_j \left( \sum_{k=1}^{k_j} \pi_{j,k} (f_{j,k}\cdot (i\nabla) + (i\nabla) \cdot f_{j,k} )\pi_{j,k} \right) \eta_j  \right] \enspace .
\end{equation}

Following \cite{gn} p.217-218 (Lemma 3.10), we have that:
\begin{lem}\label{aq} The operator $A_{M,\eta}$ is essentially self-adjoint on $C^{\infty}({\mathbb T}^d)$. For $s=1,2$, $A_{M,\eta}^s$ is relatively bounded w.r.t. $(-\Delta+1)^{s/2}$.
\end{lem}
In the following we abuse notation and denote by $A_{M,\eta}$ the self-adjoint extension of the operator defined in (\ref{conjugate2}) and $A={\cal F} A_{M,\eta} {\cal F}^{-1}$. In particular, for $s=1,2$, $A^s$ is relatively bounded w.r.t. $\bra J\ket^s$, where $J=\sum_{j\in\mathbb Z^d}j |j\ket\bra j|=- {\mathcal F} i\nabla  {\mathcal F}^{-1}$.

\begin{prop}\label{c1} Let $N\in C^1({\mathbb T}^d; {\mathcal M}_{d'}({\mathbb C}))$ such that: for all $j\in \{1,\ldots,N\}$, there exists maps  $\mu_{j,k}: \Theta_j\rightarrow {\mathbb C}$ for all $k\in \{1,\ldots,k_j\}$ and such that $\forall x\in \Theta_j,$
\begin{equation*}
N(x)= \sum_{k=1}^{k_j} \mu_{j,k}(x) \pi_{j,k}(x).
\end{equation*}
Denoting by $N$ the associated multiplication operator on $L^2({\mathbb T}^d; {\mathbb C}^{d'})$, we have that $N\in C^1(A_{M,\eta})$ and
\begin{equation*}
\mathrm{ad}_{A_{M,\eta}} N= i\sum_{j=1}^{N}\eta_j f_{j,k}\cdot(\nabla \mu_{j,k}) \pi_{j,k}.
\end{equation*}
\end{prop}
\noindent{\bf Proof:} Using sequilinear forms on $C^{\infty}({\mathbb T}^d)$, we observe first that for all $j\in \{1,\ldots,N\},$
\begin{align*}
& [\eta_j ( \sum_{k=1}^{k_j} \pi_{j,k} (f_{j,k}\cdot (i\nabla) + (i\nabla) \cdot f_{j,k}) \pi_{j,k} \eta_j, N] \\
&= \eta_j [\sum_{k=1}^{k_j} \pi_{j,k} (f_{j,k}\cdot (i\nabla) + (i\nabla) \cdot f_{j,k}) \pi_{j,k}, \sum_{k'=1}^{k_j} \mu_{j,k'} \pi_{j,k'}]\eta_j\\
&= \eta_j \sum_{k=1}^{k_j} [\pi_{j,k} (f_{j,k}\cdot (i\nabla) + (i\nabla) \cdot f_{j,k}) \pi_{j,k},\mu_{j,k}\pi_{j,k}]\eta_j\\
&= 2\eta_j \sum_{k=1}^{k_j} \pi_{j,k} (f_{j,k}\cdot (i\nabla \mu_{j,k})) \eta_j, 
\end{align*}
since for all $k\in \{1,\ldots,k_j\}$, $\pi_{j,k} (\nabla \pi_{j,k}) \pi_{j,k}= 0$. We note that the functions $\eta_j f_{j,k}\cdot (\nabla \mu_{j,k}) \pi_{j,k}$ are continuous with compact support on $\Theta$ hence uniformly bounded on ${\mathbb T}^d$. This implies that the RHS extends continuously to $L^2({\mathbb T}^d; {\mathbb C}^{d'})\times L^2({\mathbb T}^d; {\mathbb C}^{d'})$. Since $C^{\infty}({\mathbb T}^d)$ is a core for $A_{M,\eta}$, $N\in C^1(A_{M,\eta})$. The conclusion follows. \ep

We deduce that:
\begin{cor}\label{free} For $M\in C^3({\mathbb T}^d; U(d'))$, we have that $M\in C^2(A_{M,\eta})$ and
\begin{equation*}
(M^*A_{M,\eta} M- A_{M,\eta})=M^* \mathrm{ad}_{A_{M,\eta}} M=\sum_{j=1}^{N}\eta_j |\nabla \lambda_{j,k}|^2 \pi_{j,k} \enspace .
\end{equation*}
In particular, for $\Delta$  $M$-good, and $\chi_{\Delta}$ the characteristic function of $\Delta$,
\begin{equation*}
\chi_{\Delta}(M) (M^*A_{M,\eta} M- A_{M,\eta})\chi_{\Delta}(M) \geq c_{\Delta} \chi_{\Delta}(M)
\end{equation*}
where $c_{\Delta}= \min_j \min_{k\in \{1,\ldots,k_j\}} \min_{x\in K_j} |\nabla \lambda_{j,k}(x)|^2>0$. 
\end{cor}
Note that $c_{\Delta}>0$ comes from the fact that $\Delta$ is $M$-good. As a straightforward consequence of Corollary \ref{free}, we have that $\widehat{M}\in C^2(A)$ and that: 
\begin{equation*}
\chi_{\Delta}(\widehat{M}) (\widehat{M}^*A \widehat{M}- A)\chi_{\Delta}(\widehat{M}) \geq c_{\Delta} \chi_{\Delta}(\widehat{M}) \enspace .
\end{equation*}

It remains to consider the regularity conditions on the perturbation. 
The following result provides a criterion to deal with fractional regularity properties:
\begin{thm}\label{bdms} Let $Q$ be a strictly positive self-adjoint operator such that  $A^2 Q^{-2}$  is bounded.
A bounded symmetric operator $T$ belongs to ${\cal C}^{1,1}(A)$ if there exists a function $\chi \in C_0^{\infty}({\mathbb R})$ with $\chi(x)>0$ for $0<a<x<b<\infty$ such that:
\begin{equation*}
\int_1^{\infty} \left\|\chi\left( {Q}/{r}\right)T\right\| \, dr < \infty \enspace .
\end{equation*}
\end{thm}
For a proof, see Theorem 7.5.8 in \cite{abmg} or Theorem 6.1 in \cite{sah}.

\begin{lem}\label{regularsah} Let $C(j) \in U(d')$, $j\in {{\mathbb Z}^d}$,  be such that $\bf C$ is regular, i.e.  
\begin{equation}\label{decay}
\int_1^{\infty} \sup_{a r\leq |j|\leq b r} \|C(j)- 1\|\, dr < \infty \enspace ,
\end{equation}
for some $0<a<b<\infty$. Then, ${\bf C}$ and $U=\widehat M{\bf C}$ belong ${\cal C}^{1,1}(A)$. Moreover, $\mathrm{ad}_{A}({\bf C})$ is compact.
\end{lem}
\noindent{\bf Proof:} 
Let $\chi$ a smoothed characteristic function supported on (a,b) (say it takes value 1 on $[c,d]$ with $a<c<d<b$). Then, we have that:
\begin{equation*}
\int_1^{\infty}\|\chi(\bra J\ket/r)(\bf{C}-1)\|\, dr \leq \int_1^{\infty}\|1_{[a,b]}(\bra J\ket/r)(\bf{C}-1)\|\, dr \leq \int_1^{\infty}\sup_{ar\leq |j|\leq br}\|C(j)-1\|\, dr <\infty
\end{equation*}

The operator $A^2\bra J \ket ^{-2}$ is bounded, see Lemma \ref{aq}. We have that for any smoothed characteristic function $\chi$ supported on $]0,\infty[$
\begin{equation*}
\int_1^{\infty} \|\chi \left({\bra J\ket}/{r}\right)({\bf C}-1)\|\, dr < \infty \enspace .
\end{equation*}
We observe that $[\bra J\ket,{\bf C}]=0$, so
\begin{equation*}
\int_1^{\infty} \|\chi \left({\bra J\ket}/{r}\right)({\bf C}^*-1)\|\, dr < \infty \enspace .
\end{equation*}
As an application of Theorem \ref{bdms}, we deduce that $\Re ({\bf C}-1)$ and $\Im ({\bf C}-1)$ belongs to ${\cal C}^{1,1}(A)$. Since $ {\cal C}^{1,1}(A)$ is an algebra, ${\bf C}\in {\cal C}^{1,1}(A)$. Now, $\widehat{M}\in C^2(A)\subset {\cal C}^{1,1}(A)$, so that $U\in  {\cal C}^{1,1}(A)$. For the last point we refer to remark (ii) made in the proof of \cite{abmg} Theorem 7.2.9. which states that if a compact operator $B$ belongs to ${\cal C}^{1,1}(A)$, then $\mathrm{ad}_A B$ is also compact. This follows from the fact that $\mathrm{ad}_A B$ can be expressed as the norm-limit when $\tau\rightarrow 0$ of the compact operators $\tau^{-1}(e^{iA\tau}B e^{-iA\tau}-B)$, see  inclusions (5.2.10) in \cite{abmg}. Applied to ${\bf C}-1$, this yields $\mathrm{ad}_{A}({\bf C})$ is compact.
 \ep

So, combining Lemma \ref{prop-compact}, \ref{regularsah} and Corollary \ref{free} we get:
\begin{prop}\label{perturbed} Let $U=\widehat M {\bf C}$ on $l^2({\mathbb Z}^d;{\mathbb C}^{d'})$, where $ {\bf C}$ satisfies 
condition (\ref{decay}).
Then, for any real-valued continuous function $\phi\in C^0({\mathbb T})$ supported on a $M$-good set $\Delta$ 
\begin{equation*}
\phi(U) (U^* AU-A)\phi(U) \geq c_{\Lambda} \phi(U)^2 +K
\end{equation*}
where $c_{\Lambda}= \min_j \min_{k\in \{1,\ldots,k_j\}} \min_{x\in K_j} |\nabla \lambda_{j,k}|^2 >0$ and $K$ is compact. 
\end{prop}

\noindent{\bf End of the proof of Theorem \ref{mainmourre}:} This is a combination of Proposition \ref{virial}, Theorem \ref{nosc}, Proposition \ref{perturbed}, Lemma \ref{regularsah} and  Remark \ref{38}. In the analytic case, any open set $\Delta$ such that $\overline{\Delta}\subset  p_{\sigma}(\Sigma)\setminus \tau_M$ is $M$-good \cite{gn}. This implies the last statement. \ep.

\section{Applications}\label{app}

We make explicit the spectral consequences of our analysis for the unitary network models introduced in Section \ref{uml}.

\subsection{One dimensional QW}
{
\begin{prop}\label{corqw}
Let $U=S{\bf C}$ on $\bC^2\otimes l^2(\bZ;\bC)$ be a symmetric one dimensional quantum walk 
\[{\bf C}=\sum_{j\in\bZ}C(j)C_\infty\otimes|j\ket\bra j|\]
with $C(j),C_\infty\in U(2)$ and the family $C(j)$ satisfying the regularity condition (\ref{eq:regularity}). 
Let 
\[
C_{\infty}= e^{-i\eta} \left( \begin{array}{cc} \alpha & -\bar{\beta} \\ \beta & \bar{\alpha} \end{array} \right)\qquad \hbox{ for } \eta\in\bT, \alpha, \beta\in\bC, \vert\alpha\vert^2+\vert\beta\vert^2=1. 
\]

Then it holds for the spectrum of $U$:
\[ \sigma_{sc}(U)= \emptyset, \ \ \sigma_{ess}(U)=B_+\bigcup B_{-}, \]
with the bands
\[ B_{\pm}:=\left\{ e^{-i \eta}\left(p\pm i\sqrt{1-p^2}\right),p\in\lbrack-\vert\alpha\vert,\vert\alpha\vert\rbrack\right\}.\]
Furthermore
$\tau_M=\left\{ e^{-i \eta}\left(\vert\alpha\vert\pm i\sqrt{1-\vert\alpha\vert^2}\right),e^{-i \eta}\left(-\vert\alpha\vert\pm i\sqrt{1-\vert\alpha\vert^2}\right)\right\}.$
$$\sigma_{ess}(U)\setminus \tau_M\subset\sigma_{ac}(U)\cup \sigma_{d}(U)\quad\hbox{ if }\alpha\neq0$$
 and discrete eigenvalues can only accumulate at $\tau_M$. \\
 If $\vert\alpha\vert=0$, $U$ is pure point and $\sigma_{ess}(U)=\left\{\pm i e^{-i\eta}\right\}$.
\end{prop}
{\bf Proof.} The matrix valued multiplication operator $M$ defined in (\ref{mqw}) is defined by
\[M(x)= \left( \begin{array}{cc}e^{ix} & 0\\ 0& e^{-ix} \end{array} \right)e^{-i\eta} \left( \begin{array}{cc} \alpha & -\bar{\beta} \\ \beta & \bar{\alpha} \end{array} \right)\qquad \hbox{ for } x\in\bT. \]
The spectrum of $M(x)$ as a set are the values of  

\[\lambda_{\pm}(x)=e^{-i \eta}\left(p(x)\pm i\sqrt{1-p(x)^2}\right)\] 
where
\[ p(x):=\vert\alpha\vert\cos(\varphi(x))\qquad(x\in\mathbb T)\]
and $\varphi\in C^1(\bT)$ satisfies $e^{i\varphi(x)}=\frac{\alpha}{\vert\alpha\vert}e^{ix}$.
The function $p$ is onto  $\lbrack-\vert\alpha\vert,\vert\alpha\vert\rbrack$ and thus 

$\lambda_\pm(\bT)=\left\{e^{-i\eta}\left(p\pm i\sqrt{1-p^2}\right), p\in\lbrack-\vert\alpha\vert,\vert\alpha\vert\right\}$.

Band crossings $\lambda_{+}(x)=\lambda_{-}(x)$ occur only in the case $\vert\alpha\vert=1$ and $\vert p(x)\vert=1$, i.e. at quasienergies
$
\left\{\pm e^{-i\eta}\right\}.
$
Critical points $\nabla\lambda_{\pm}(x)=0$ occur at the critical points of $p$, i.e. if $\sin(\varphi(x))=0$.
We conclude that the quasienergy values which correspond to bandcrossings and critical points occur at all band edges i.e. 
\[  \tau_M:=\left\{e^{-i\eta}\left(\vert\alpha\vert\pm i\sqrt{1-\vert\alpha\vert^2}\right),e^{-i\eta}\left(-\vert\alpha\vert\pm i\sqrt{1-\vert\alpha\vert^2}\right)\right\}\]
and that any spectral interval $\Delta$ which avoids these points is $M$-good in the sense of Theorem 3.3 from which our claim follows.\ep

\subsection{BB Matrices}

Consider now BB matrices which have a slightly different structure. 
\begin{cor}\label{bbscat}
Let $U_{\text{BB}}$ be constructed from $\{S_k\}_{k\in\mathbb Z}$ given by (\ref{scatt}), and let $\Sigma=\begin{pmatrix} i & 0\\ 0& -i\end{pmatrix}$. Suppose there exists $S_{\infty}= e^{-i\eta} \left( \begin{array}{cc} \alpha & -\bar{\beta} \\ \beta & \bar{\alpha} \end{array} \right) \in U(2)$ such that ${\bf C}$ defined via
\begin{equation*}
C(k)=\left\{\begin{matrix} S_\infty^{-1}S_k  & \mbox{if} \ k \ \mbox{even} \\
S_\infty^{-1}\Sigma S_k\Sigma &\mbox{if } \ k \ \mbox{odd}\end{matrix}\right. \ \ \ \mbox{is regular.}
\end{equation*}
Then, upon replacing  $\lambda_\pm$ by $\lambda_\pm^2$ and $\tau_M$ by $\tau_M^2$, the spectral conclusions of Proposition \ref{corqw} apply to $\sigma(U_{\text{BB}})$.
\end{cor}
\begin{rem} This Corollary generalizes Theorem 6.2 of \cite{bhj} on absence of singular continuous spectrum for BB matrices.
Note, however, that transfer matrix methods allow to say much more on the spectrum in one dimension, see \cite{bhj, dfv}.
\end{rem}
{\bf Proof:} The result follows from  Proposition \ref{proot}, Proposition \ref{corqw} and spectral mapping.\ep

\subsection{Chalker Coddington Model}

\begin{prop}
Let ${\bf D}:=\sum_{j\in\bZ^2}D(j)\otimes |j\ket\bra j|$ on $\bC^4\otimes l^2(\bZ^2;\bC)$ with $D(j)$ unitary and diagonal be regular in the sense of Definition \ref{regular}. Consider $\varphi\in\lbrack0,\frac{\pi}{2}\rbrack$ and the $U(\varphi)$  defined in (\ref{def:u}).
Then it holds for the spectrum of $U(\varphi)$:
\[ \sigma_{sc}(U(\varphi))= \emptyset, \ \ \sigma_{ess}(U(\varphi))=\bigcup_{j,k\in\{+,-\}}B_{j,k}, \]
with the bands
\[ B_{+,\pm}:=\left\{\pm e^{i x}\in {\partial \mathbb D}; x\in\lbrack-\varphi,\varphi\rbrack\right\}, \quad B_{-,\pm}:=\left\{\pm i e^{i x}\in {\partial \mathbb D}; x\in\lbrack-\varphi,\varphi\rbrack\right\}.\]
Furthermore, for $\varphi\neq 0$, 
$\tau_M=\left\{ +e^{\pm i \varphi}, -e^{\pm i \varphi}, i e^{\pm i\varphi}, - i e^{\pm i\varphi}, \pm 1, \pm i\right\}$,
$$\sigma_{ess}(U(\varphi))\setminus \tau_M\subset\sigma_{ac}(U(\varphi))\cup \sigma_{d}(U(\varphi)),$$
 and discrete eigenvalues can only accumulate at $\tau_M$. \\
 If $\varphi=0$, $U(0)$ is pure point and $\sigma_{ess}(U(0))=\left\{\pm 1, \pm i\right\}$.
\end{prop}
{\bf Proof.} The operator ${\bf D}^{-1}U(\varphi)$ is represented by the matrix valued multiplication operator $M$ defined in (\ref{mcc}). The spectrum of $M(x,y)$ as a set are the square roots $\lambda_{+,\pm}=+\sqrt{\mu_\pm}, \lambda_{-,\pm}=-\sqrt{\mu_\pm}$ of 
\[\mu_\pm=i p\pm\sqrt{1-p^2}\]
where
\[ p(x,y):=\sin{(2\varphi)}h(x,y), \quad h(x,y):=\frac{1}{2}(\cos x+\cos y)\qquad(x,y)\in\mathbb T^2.\]
The function $h$ is onto on $\lbrack-1,1\rbrack$ thus $\mu_\pm(\mathbb T^2)=\left\{\pm e^{ix}; x\in\lbrack-2\varphi,2\varphi\rbrack\right\}$ which implies the band structure of the spectrum of $M$.

Band crossings $\lambda_{jk}(x,y)=\lambda_{lm}(x,y)$ occur only in the case $\varphi=\frac{\pi}{4}$ and $\vert p(x,y)\vert=1$, i.e. at quasienergies
\[
\left\{e^{\pm i\varphi}, -e^{\pm i\varphi}\right\}_{\varphi=\frac{\pi}{4}}\hbox{  and \ } (x,y)\in\left\{(0,0),(\pi,\pi)\right\}.
\]
Critical points $\nabla\lambda_{jk}(x,y)=0$ occur at the critical points of $h$, i.e. at 
\[(x,y)\in\left\{(0,0),(\pi,0), (0,\pi), (\pi,\pi)\right\}.\]
We conclude that the quasienergy values which correspond to bandcrossings and critical points occur at all band edges and all band centers, i.e. 
\[  \tau_M:=\left\{e^{\pm i\varphi}, -e^{\pm i\varphi}, i e^{\pm i \varphi},  -i e^{\pm i \varphi}, \pm 1, \pm i\right\}\]
and that any spectral interval $\Delta$ which avoids these points is $M$-good in the sense of Theorem 3.3 from which our claim follows.\ep

\subsection{Symmetric QW} 
For a generic $d$-dimensional symmetric quantum walk, the analytic matrix (\ref{mqw}) representing $U_\infty$ cannot be diagonalized explicitly, which prevents us from determining exactly the discrete set $\tau_M$. We nevertheless get from Theorem \ref{mainmourre}
\begin{cor}\label{corddqw}
Let  $U_\infty$ on $L^2(\mathbb Z^d;\mathbb C^{2d})$ be represented in Fourier space by the multiplication operator by $M(x)$, $x\in\mathbb T^d$, given by (\ref{mqw}), and $U:=U_\infty {\bf C}$, with ${\bf C}$ regular, as in Section \ref{QW}. Then
\begin{equation*}
\sigma_{sc}(U)=\emptyset, \ \ \sigma_{ess}(U)=\cup_{x\in\mathbb T^d} \sigma(M(x)), \ \ \sigma_{ess}(U)\setminus\tau_M\subset\sigma_{ac}(U)\cup \sigma_{d}(U), 
\end{equation*}
and the finitely degenerate eigenvalues of $U$ can accumulate at  $\tau_M$ only. 
 \end{cor}
\begin{rem} As the one dimensional case shows, $\sigma_{ac}(U)$ or $\sigma_{d}(U)$ can be empty.
\end{rem}

In case the homogeneous quantum walk $U_\infty$ is given by a direct sum of decoupled one dimensional quantum walks, {\it i.e.}, $C_\infty=\oplus_{k=1}^d C_\infty(k)\in U(2d)$, with $C_\infty(k)\in U(2)$, we have $\tau_M=\cup_{k=1}^d \tau_M(k)$, where $\tau_M(k)$ is described in Proposition \ref{corqw}, for $k=1,\dots,d$. In such a case, the perturbed operator $U=U_\infty{\bf C}$, with  ${\bf C}$ regular, generically describes $d$ coupled one dimensional quantum walks and its spectrum is characterized by Proposition \ref{corqw} and Corollary \ref{corddqw}.


\end{document}